\documentclass[11pt,a4paper]{article}
\pdfoutput=1
\usepackage{jcappub}
\usepackage[T1]{fontenc}
\usepackage{color}
\usepackage{graphicx}
\usepackage{bm}
\usepackage{amsmath,amsfonts,amssymb}
\usepackage{braket}

\begin{document}


\title{Lepton Asymmetric Universe}
\author[a,b]{Masahiro Kawasaki}
\author[a,b]{Kai Murai}
\affiliation[a]{ICRR, University of Tokyo, Kashiwa, 277-8582, Japan}
\affiliation[b]{Kavli IPMU (WPI), UTIAS, University of Tokyo, Kashiwa, 277-8583, Japan}

\abstract{
The recent observation of $^4$He implies that our universe has a large lepton asymmetry.
We consider the Affleck-Dine (AD) mechanism for lepton number generation.
In the AD mechanism, non-topological solitons called L-balls are produced, and the generated lepton number is confined in them.
The L-balls protect the generated lepton number from being converted to baryon number through the sphaleron processes.
We study the formation and evolution of the L-balls and find that the universe with large lepton asymmetry suggested by the recent $^4$He measurement can be realized.
}

\keywords{
physics of the early universe, leptogenesis, supersymmetry and cosmology, big bang nucleosynthesis
}

\emailAdd{kawasaki@icrr.u-tokyo.ac.jp}
\emailAdd{kmurai@icrr.u-tokyo.ac.jp}

\maketitle

\section{Introduction}
\label{sec: intro}

Recently, Matsumoto \textsl{et al.}~\cite{Matsumoto:2022tlr} observed 10 extremely metal-poor ($Z < 0.1 Z_\odot$) galaxies and measured their $^4$He abundances.
Together with the previously observed 54 galaxies, they determined the primordial $^4$He abundance as
\begin{align}
    Y_p = 0.2370^{+0.0034}_{-0.0033}
\end{align}
which is smaller than the previous measurements~\cite{Hsyu:2020uqb,Aver:2015iza,Izotov:2014fga}.
Furthermore, using the recent deuterium (D) measurement~\cite{Cooke:2017cwo} and the baryon-to-photon ratio $\eta_b$ from Planck 2018~\cite{Planck:2018vyg}, Ref.~\cite{Matsumoto:2022tlr} obtained the constraints on the degeneracy parameter of electron neutrinos $\xi_e$ and the effective number of neutrino species $N_\mathrm{eff}$ as
\begin{align}
   \xi_e  &  = 0.05^{+0.03}_{-0.02},
   \label{eq:chemical_pot}
   \\[0.5em]
   N_\mathrm{eff} & = 3.11^{+0.34}_{-0.31}.
\end{align}
The nonzero $\xi_e$ implies that our universe has a lepton asymmetry.
From Eq.~\eqref{eq:chemical_pot}, the lepton asymmetry of the universe is estimated as $\sim 5 \times 10^{-3}$, which is much larger than the baryon asymmetry $\sim 10^{-10}$.
Such large  positive lepton asymmetry and $N_\mathrm{eff}$ are also favored in the context of the Hubble tension~\cite{Matsumoto:2022tlr,Seto:2021tad}.

Lepton number is generated through some processes like right-handed neutrino decay~\cite{Fukugita:1986hr} and Affleck-Dine mechanism~\cite{Affleck:1984fy,Dine:1995kz}.
However, the lepton number produced above the electroweak scale $T\gtrsim 100$~GeV is partially converted to baryon number through the sphaleron processes.
As a result, the produced lepton and baryon asymmetries become of the same order. 
Therefore, in order to produce a lepton asymmetry much larger than the baryon asymmetry, lepton number should be produced below the electroweak scale or the produced lepton asymmetry should be protected against the sphaleron processes. 
The former can be realized in the right-handed neutrino decay.
However, the maximum lepton asymmetry in this model is $\sim 7\times 10^{-4}$~\cite{Boyarsky:2009ix}, which is smaller than that required from Eq.~\eqref{eq:chemical_pot}.
The Q-ball formation associated with the Affleck-Dine leptogenesis belongs to the latter.

The Q-ball is a non-topological soliton in scalar field theories with a global U(1) symmetry~\cite{Coleman:1985ki}. 
It is known that Q-balls are produced in the Affleck-Dine mechanism~\cite{Kusenko:1997si,Enqvist:1997si,Kasuya:1999wu}, which utilizes flat directions in the minimal supersymmetric standard model (MSSM). 
When some flat direction with lepton number (not baryon number) has a large field value during inflation, it generates lepton number by field dynamics after inflation. 
At the same time, Q-balls are produced and the generated lepton number is all confined inside the Q-balls. 
Thus, such Q-balls have a large lepton number and are called L-balls.
Since the sphaleron effect is inactive inside the L-balls, the lepton asymmetry produced by the Affleck-Dine mechanism are kept without being converted to a baryon asymmetry~\cite{Kawasaki:2002hq,Gelmini:2020ekg}. 
The L-balls decay into neutrinos below the electroweak scale and realize a large lepton asymmetry in the universe. 

In this paper, we consider the formation and evolution of the L-balls and study whether a large lepton asymmetry is produced.
The L-ball scenario was studied by one of the authors and collaborators~\cite{Kawasaki:2002hq} (see also \cite{Gelmini:2020ekg}). 
We update the analysis with improved treatment of the L-ball properties (e.g., mass, radius)~\cite{Hisano:2001dr}, decay~\cite{Kawasaki:2012gk}, and evaporation~\cite{Kasuya:2014ofa} and find that the universe with large lepton asymmetry suggested by the recent $^4$He measurement is realized.

This paper is organized as follows.
In Section~\ref{sec: BBN and lepton symmetry}, we briefly review the effects of lepton asymmetry on the big bang nucleosynthesis (BBN) and quantify the lepton asymmetry required to explain the new determination of the primordial $^4$He abundance.
We describe the properties of L-balls and discuss the generated lepton asymmetry and the cosmological constraints on the L-ball scenario in Section~\ref{sec: l-ball}.
Section~\ref{sec: summary and discussion} is devoted to the summary and discussion of our results.

\section{Big bang nucleosynthesis and lepton asymmetry}
\label{sec: BBN and lepton symmetry}

In the big bang cosmology, the light elements, D, $^3$He, $^4$He, and $^7$Li, are considered to be synthesized below the temperature $T \sim 1$~MeV.
The theoretical predictions of the abundances of D and $^4$He are in good agreement with the estimation of the astronomical observations.
Then, BBN provides a constraint on scenarios that spoil this agreement.
In particular, the primordial $^4$He abundance is mostly determined by the freeze-out of the neutron-proton ratio, which depends on the Hubble parameter and the electron-neutrino distribution at $T \sim 1$~MeV.
Thus, $N_\mathrm{eff}$ and $\xi_e$ can be constrained by the observational determination of the primordial $^4$He abundance.

From the new determination of primordial $^4$He abundance, a positive asymmetry of electron neutrinos is favored:
\begin{align}
    \xi_{e} \equiv \frac{\mu_{\nu_e}}{T}
    \sim
    0.05,
\end{align}
where $\mu_{\nu_e}$ is the chemical potential of the electron neutrino.
Since
\begin{align}
    n_{\nu_e} - n_{\bar{\nu}_e}
    \simeq
    \frac{T^3}{6} \xi_{e},
\end{align}
we obtain
\begin{align}
    \frac{n_{\nu_e} - n_{\bar{\nu}_e}}{n_\gamma} 
    &\simeq
    \frac{T^3 \xi_{e}/6}{2 \zeta(3) T^3/\pi^2}
    \simeq
    0.68 \xi_{e}
    \simeq
    3.4 \times 10^{-2},
\\
    \eta_{\nu_e}
    \equiv
    \frac{n_{\nu_e} - n_{\bar{\nu}_e}}{s} 
    &\simeq
    \frac{T^3 \xi_{e}/6}{2 \pi^2 g_{*} T^3/45}
    \simeq
    0.035 \xi_{e}
    \simeq
    1.75 \times 10^{-3},
\end{align}
where $g_*$ is the relativistic degree of freedom and we used $g_* = 10.75$ as the value at the BBN epoch.
Since we can consider that the asymmetries in the three flavors of neutrinos are the same due to the neutrino oscillation, the new measurement of $^4$He suggests
\begin{align}
    \eta_L 
    \equiv
    \frac{n_L}{s}
    =
    3 \eta_{\nu_e}
    \simeq
    5.3 \times 10^{-3},
\end{align}
where $n_L$ is the total lepton asymmetry.

\section{Lepton asymmetry from L-ball}
\label{sec: l-ball}

In this section, we present a model for generating a large lepton asymmetry without overproduction of a baryon asymmetry.
The model is based on the Affleck-Dine mechanism in the minimal supersymmetric standard model (MSSM)~\cite{Affleck:1984fy,Dine:1995kz}. 
In the MSSM, there exist many flat directions in the scalar potential of squark, slepton and Higgs fields~\cite{Gherghetta:1995dv} and some flat directions can have large field values during inflation.
Such flat directions start to oscillate after inflation and produce a baryon/lepton asymmetry if they have a baryon/lepton charge.
This is called Affleck-Dine mechanism for baryogenesis/leptogenesis (for a review, see Ref.~\cite{Dine:2003ax}).

Here, we focus on the Affleck-Dine leptogenesis in gauge-mediated SUSY breaking scenario and suppose that some flat direction (called AD field $\phi$) with a lepton charge has a large field value during inflation.
The potential of the AD field for $|\phi| \gg M_m$ is written as
\begin{align}
    V(\phi) 
    &= 
    V_\mathrm{gauge} + V_\mathrm{grav} + V_A
\nonumber\\
    &= 
    M_F^4 \left[\log\left(\frac{|\phi|^2}{M_m^2}\right)\right]^2 
    + m_{3/2}^2 |\phi|^2 \left(1+ K \log\frac{|\phi|^2}{M_*^2}\right)
    + V_A,
\end{align}
where $M_m$ is the messenger scale, $M_F$ is the SUSY breaking scale, $m_{3/2}$ is the gravitino mass, $K$ is the coefficient of the one-loop correction, and $M_*$ is the renormalization scale.
Here, $V_A$ is the A-term which generates the lepton asymmetry.
$V_\mathrm{gauge}$ denotes the potential coming from the gauge-mediated SUSY breaking~\cite{deGouvea:1997afu}, and $V_\mathrm{grav}$ is due to the gravity mediation. 
For $|\phi| \gtrsim \varphi_\mathrm{eq} \simeq \sqrt{2} M_F^2/m_{3/2}$, the potential is dominated by $V_\mathrm{grav}$.
We assume that the field value $\varphi \equiv |\phi|$ during inflation is much larger than $\varphi_\mathrm{eq}$.
When the Hubble parameter $H$ becomes comparable to the effective mass, i.e. $3 H\simeq m_{3/2}$, the AD field starts to oscillate.
At the same time, the AD field is kicked in the phase direction due to the A-term $V_A$, which leads to the generation of a lepton asymmetry.
The produced lepton asymmetry is given by
\begin{equation}
    n_L 
    \simeq
    \varepsilon m_{3/2}\varphi_\mathrm{osc}^2
    \label{eq: AD lepton asymmetry}
\end{equation}
where $\varphi_\mathrm{osc} > \varphi_\mathrm{eq}$ is the field value at the onset of the oscillation and $\varepsilon (\le 1)$ is the parameter which represents the efficiency of the asymmetry generation due to the A-term.
In the following, we take $\varepsilon =1$ for simplicity.\footnote{%
When the field value during inflation is determined by the balance between the non-renormalizable term and the Hubble induced term~\cite{Dine:1995kz}, the effective mass in the phase direction at the start of oscillation is $\sim m_{3/2}$, which leads to $\varepsilon \sim \sin(N\theta)$, where $\theta$ is the initial phase of the AD field and $N=4$ or $6$.
}

Here and hereafter, we assume that the oscillation of the AD field starts before reheating.
Then, we can set the upper bound on the reheating temperature $T_R$ as follows.
The above assumption is written as
\begin{align}
    H_\mathrm{osc} \gtrsim H_R,
    \label{eq: oscillation before reheating}
\end{align}
where $H_\mathrm{osc}$ and $H_\mathrm{R}$ are the Hubble parameter at the onset of oscillation and reheating, respectively.
Since $3 H_\mathrm{osc} \simeq m_{3/2}$ and $M_\mathrm{Pl}^2 H_R^2 \sim T_R^4$, Eq.~\eqref{eq: oscillation before reheating} is rewritten as
\begin{align}
    T_R 
    \lesssim 
    T_{R,\mathrm{max}}
    \sim
    \sqrt{m_{3/2} M_\mathrm{Pl}}
    \sim
    10^9~\mathrm{GeV} \, \left( \frac{ m_{3/2} }{0.5~\mathrm{GeV} } \right)^{1/2}.
    \label{eq: TR upper bound}
\end{align}
If this condition is not satisfied, the AD field starts to oscillate at $T_\mathrm{osc} = T_{R,\mathrm{max}}$ and the following estimations are valid with a replacement of $T_R$ by $T_\mathrm{osc}$.

Another important consequence of the AD mechanism is the formation of Q-balls~\cite{Coleman:1985ki,Kusenko:1997zq,Dvali:1997qv,Kusenko:1997si,Enqvist:1997si,Kasuya:1999wu}. 
A Q-ball is a non-topological soliton made of a complex scalar field, which is stable due to a global U(1) symmetry.
During oscillation, the AD field fragments into spherical lumps through spatial instabilities of the field and almost all lepton number is confined within Q-balls. 
When the potential is dominated by $V_\mathrm{grav}$ or equivalently $|\phi| \gtrsim \varphi_\mathrm{eq}$, the Q-ball formation depends on whether $K$ is positive or negative.
For $K<0$, the Q-balls are formed soon after the AD field starts to oscillate.
On the other hand, the Q-ball formation is delayed until the amplitude of the AD field becomes $\varphi_\text{eq}$ for $K > 0$.
This is because $V_\text{grav}$ with $K > 0 (<0)$ does (not) allow a Q-ball solution while $V_\text{gauge}$ always has a Q-ball solution.
This type of Q-balls is called delayed-type Q-balls~\cite{Kasuya:2001hg}.

\subsection{Lepton asymmetry from delayed-type L-balls}
\label{subsec: delayed type}

In the following, we assume $K > 0$ and focus on delayed-type Q-balls with lepton charge, i.e., delayed-type L-balls.
The reason why the delayed-type L-ball is favored is as follows.
In order to generate a large lepton asymmetry, the AD field should have a large field value at the onset of oscillation, which leads to the formation of L-balls with a large lepton charge if the L-ball formation takes place at the same time.
Such large L-balls have a lifetime too long to decay before BBN.
On the other hand, the delayed-type L-balls have a smaller lepton charge and hence they can decay before BBN.

The properties of a delayed-type L-ball are given by~\cite{Hisano:2001dr}
\begin{align}
\begin{aligned}
    Q
    &=
    \beta \left( \frac{ \varphi_\mathrm{eq} }{ M_F } \right)^4,
\\
    M_Q
    &=
    \frac{4 \sqrt{2} \pi}{3} \zeta M_F Q^{3/4},
\\
    R_Q 
    &= 
    \frac{1}{\sqrt{2} \zeta} M_F^{-1} Q^{1/4},
\\
    \omega_Q
    &\simeq
    \sqrt{2} \pi \zeta M_F Q^{-1/4},
\end{aligned}
    \label{eq: Q-ball formula for delayed}
\end{align}
where $Q$ is the L-ball charge or lepton number, $M_Q$ is the L-ball mass, $R_Q$ is the L-ball radius, and $\omega_Q$ is the energy per charge of the L-ball.
$\beta \simeq 6 \times 10^{-4}$ is a dimensionless constant~\cite{Kasuya:2001hg}.

L-balls decay by emitting neutrinos with the decay rate~\cite{Cohen:1986ct,Kawasaki:2012gk},
\begin{equation}
    \Gamma_Q 
    \simeq
    \frac{N_\ell}{Q} \frac{\omega_Q^3}{12\pi^2} 4\pi R_Q^2,
\end{equation}
where $N_\ell$ is the number of decay channels.
Then, the decay rate of the delayed-type L-balls is given by
\begin{align}
    \Gamma_Q
    &\simeq
    \frac{\pi^2 N_\ell \zeta}{12 \beta^{5/4}} 
    \frac{m_{3/2}^5}{M_F^4},
    \label{eq: decay rate formula for delayed}
\end{align}
where we used Eq.~\eqref{eq: Q-ball formula for delayed} and $\varphi_\mathrm{eq} \simeq \sqrt{2} M_F^2/m_{3/2}$.

Next, we estimate the cosmic temperature $T_\mathrm{D}$ when the L-balls decay.
$T_\mathrm{D}$ should be higher than $\sim \mathcal{O}(\mathrm{MeV})$ in order to affect the BBN through the generated lepton asymmetry.
$T_\mathrm{D}$ is estimated as
\begin{align}
    T_\mathrm{D} 
    & \simeq 
    \left( \frac{90}{\pi^2 g_*(T_\mathrm{D})} \right)^{1/4}
    \sqrt{M_\mathrm{Pl}\Gamma_{Q}}
\nonumber\\
    &=
    \left( \frac{90}{\pi^2 g_*(T_\mathrm{D})} \right)^{1/4}
    \frac{\pi N_\ell^{1/2} \zeta^{1/2}}{2\sqrt{3} \beta^{5/8}} 
    \frac{M_\mathrm{Pl}^{1/2} m_{3/2}^{5/2}}{M_F^2}
\nonumber\\
    &\simeq
    2.69 ~\mathrm{MeV} \,
\nonumber\\
    & \hspace{15pt} \times 
    \left( \frac{g_*}{10.75} \right)^{-1/4}
    \left( \frac{\beta}{6 \times 10^{-4}} \right)^{-5/8}
    \left( \frac{m_{3/2}}{0.5~\mathrm{GeV}} \right)^{5/2}
    \left( \frac{M_F}{5\times 10^6~\mathrm{GeV}} \right)^{-2}
    \left( \frac{N_\ell}{3} \right)^{1/2}
    \left( \frac{\zeta}{2.5} \right)^{1/2}
    ,
    \label{eq: decay temperature formula for delayed}
\end{align}
where $g_*(T_\mathrm{D})$ is the relativistic number of degrees of freedom at $T_\mathrm{D}$.

Next, we assume that the L-balls dominate the universe at the decay and evaluate the lepton asymmetry $\eta_L$.
In this case, the ratio of the generated lepton asymmetry $n_L$ to the energy density of the L-balls $\rho_Q$ remains constant until the L-ball decay, and the entropy density of the decay product is given by $s = 4\rho_Q|_{T_\mathrm{D}}/(3T_\mathrm{D})$ at the decay.
Therefore, we obtain
\begin{equation}
    \eta_L 
    \simeq
    \frac{m_{3/2}\varphi_\mathrm{osc}^2}{4m_{3/2}^2 \varphi_\mathrm{osc}^2/(3 T_\mathrm{D})}
    =
    \frac{3 T_\mathrm{D}}{4 m_{3/2}},
    \label{eq: lepton asymmetry for delayed Q-ball}
\end{equation}
which leads to the estimation of the decay temperature as
\begin{align}
    T_\mathrm{D}
    &=
    \frac{4}{3} \eta_L m_{3/2}
\nonumber\\
    &
    \simeq
    3.3~\mathrm{MeV} \,
    \left( \frac{m_{3/2}}{0.5~\mathrm{GeV}} \right)
    \left( \frac{\eta_L}{5 \times 10^{-3}} \right).
\end{align}
Then, if we require $T_\mathrm{D} > 1$~MeV, we obtain the condition for the BBN as
\begin{align}
    m_{3/2} > 
    0.15~\mathrm{GeV}
    \left( \frac{\eta_L}{5 \times 10^{-3}} \right)^{-1}
    .
\end{align}
On the other hand, from Eq.~\eqref{eq: decay temperature formula for delayed}, the lepton asymmetry can also be evaluated as
\begin{align}
    \eta_L
    & \simeq 
    \left( \frac{90}{\pi^2 g_*(T_\mathrm{D})} \right)^{1/4}
    \frac{\sqrt{3} \pi N_\ell^{1/2} \zeta^{1/2}}{8 \beta^{5/8}} 
    \frac{M_\mathrm{Pl}^{1/2} m_{3/2}^{3/2}}{M_F^2}
\nonumber\\
    & \simeq 
    4.04 \times 10^{-3} \,
\nonumber\\
    & \hspace{15pt} \times 
    \left( \frac{g_*}{10.75} \right)^{-1/4}
    \left( \frac{\beta}{6 \times 10^{-4}} \right)^{-5/8}
    \left( \frac{m_{3/2}}{0.5~\mathrm{GeV}} \right)^{3/2}
    \left( \frac{M_F}{5 \times  10^6~\mathrm{GeV}} \right)^{-2}
    \left( \frac{N_\ell}{3} \right)^{1/2}
    \left( \frac{\zeta}{2.5} \right)^{1/2}
    .
\end{align}

Since we assumed the L-ball domination at the decay in the above argument, we confirm that this assumption is valid for the benchmark set of the parameters.
The energy density ratio of the L-balls to radiation at $T_\mathrm{D}$ is estimated by
\begin{align}
    \left. \frac{\rho_Q}{\rho_R} \right|_{T_\mathrm{D}}
    &\simeq
    \frac{ m_{3/2}^2 \varphi_\mathrm{osc}^2 }{ 3 M_\mathrm{Pl}^2 H_\mathrm{osc}^2 }
    \frac{T_R}{T_\mathrm{D}}
\nonumber\\
    &=
    \frac{12\sqrt{3} \beta^{5/8}}{\pi N_\ell^{1/2} \zeta^{1/2}}
    \left( \frac{\pi^2 g_*(T_\mathrm{D})}{90} \right)^{1/4}
    \frac{T_R M_F^6}{M_\mathrm{Pl}^{5/2} m_{3/2}^{9/2}}
    \left( \frac{\varphi_\mathrm{osc}}{\varphi_\mathrm{eq}} \right)^2
\nonumber\\
    &\simeq
    9.66 \times 10^6 \,  
    \left( \frac{g_*}{10.75} \right)^{1/4}
    \left( \frac{\beta}{6 \times 10^{-4}} \right)^{5/8}
    \left( \frac{m_{3/2}}{0.5~\mathrm{GeV}} \right)^{-9/2}
    \left( \frac{M_F}{5 \times 10^6~\mathrm{GeV}} \right)^{6}
\nonumber\\
    & \hspace{53pt} \times
    \left( \frac{N_\ell}{3} \right)^{-1/2}
    \left( \frac{\zeta}{2.5} \right)^{-1/2}
    \left( \frac{T_R}{ 10^5~\mathrm{GeV}} \right)
    \left( \frac{\varphi_\mathrm{osc}}{10^4 \varphi_\mathrm{eq}} \right)^2
    ,
    \label{eq:entropy_prod}
\end{align}
which is larger than unity for the benchmark parameters and assure the L-ball domination.

\begin{figure}[t]
	\centering
	\includegraphics[width=.8\textwidth]{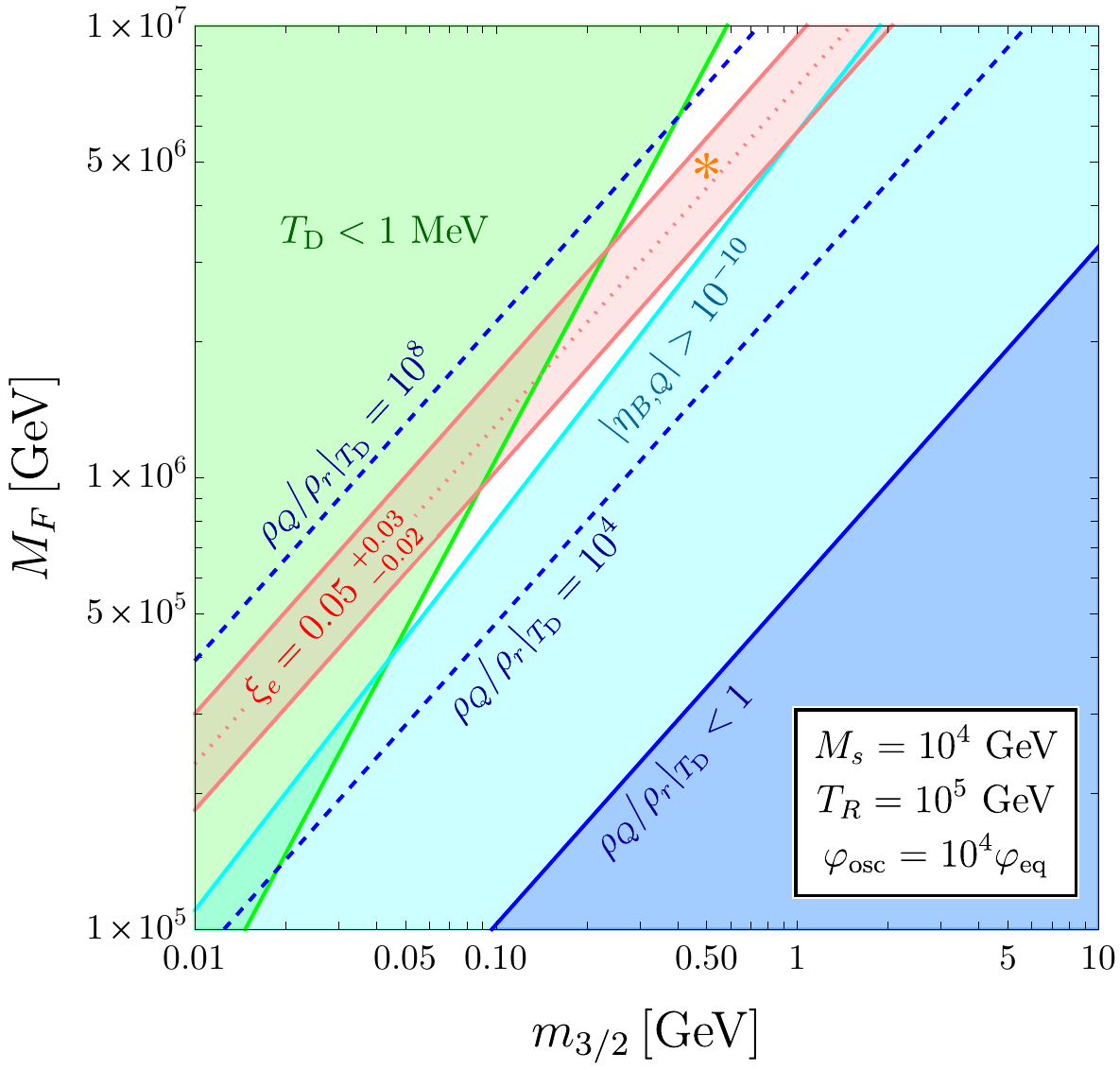}
	\caption{
	Lepton asymmetry and cosmological constraints in the L-ball scenario.
	We use $M_s = 10^4$~GeV, $T_R = 10^5$~GeV, and $\varphi_\mathrm{osc} = 10^4 \varphi_\mathrm{eq}$.
	The red shaded region corresponds to $\xi_e = 0.05^{+0.03}_{-0.02}$
	and the red dotted line shows the center value.
	Above the green line, the L-ball decay temperature becomes lower than $1$~MeV and the generated lepton asymmetry does not affect the BBN.
	Blow the blue line, the L-balls do not dominate the universe at the decay and our estimation of the lepton asymmetry becomes invalid.
	The blue dashed lines correspond to $\rho_Q/\rho_r|_{T_\mathrm{D}} = 10^4$ and $10^8$.
	Below the cyan line, since $|\eta_{B,Q}| > 10^{-10}$, the scenario needs a fine-tuning between $\eta_{B,Q}$ and the baryon asymmetry generated from some baryogenesis mechanism.
	The orange star shows the benchmark point, where $m_{3/2} = 0.5$~GeV and $M_F = 5 \times 10^6$~GeV.
	}
	\label{fig: lepton asymmetry and constraints}
\end{figure}

Here, we discuss the gravitino problem.
Gravitinos are produced at the reheating epoch after inflation.
They are stable in gauge-mediated SUSY breaking and give a significant contribution to the matter density.
When the reheating temperature $T_R$ is high, gravitinos are overproduced, which causes a cosmological difficulty called gravitino problem.
Without entropy production due to the L-ball decay, the gravitino density parameter $\Omega_{3/2}$ is given by~\cite{Kawasaki:2017bqm}
\begin{equation}
   \label{eq:gravitino_prod}
   \Omega_{3/2} h^2 \simeq 0.71\, 
   \left(\frac{m_{3/2}}{0.5\,\text{GeV}}\right)^{-1}
   \left(\frac{M_{\tilde{g}}}{10^4\,\text{GeV}}\right)^2
   \left(\frac{T_R}{10^5\,\text{GeV}}\right),
\end{equation}
where $M_{\tilde{g}}$ is the gluino mass and $h$ is the present Hubble parameter in units of 100km/s/Mpc.
Including the effect of the entropy production, Eq.~\eqref{eq:gravitino_prod} is rewritten as
\begin{equation}
   \Omega_{3/2} h^2 \simeq 0.71\,
   \left(\frac{m_{3/2}}{0.5\,\text{GeV}}\right)^{-1}
   \left(\frac{M_{\tilde{g}}}{10^4\,\text{GeV}}\right)^2
   \left(\frac{T_R}{10^5\,\text{GeV}}\right) 
   \left(\left.\frac{\rho_Q}{\rho_R}\right|_{T_\mathrm{D}}\right)^{-3/4}.
\end{equation}
This should be less than the dark matter density $\Omega_\text{DM}h^2 \simeq 0.12$, which leads to the constraint on $T_R$.
From Eq.~\eqref{eq:entropy_prod}, the constraint from the gravitino overproduction is much relaxed by the large entropy production due to the L-balls.
As we will see in Fig.~\ref{fig: lepton asymmetry and constraints}, in the parameter region explaining the favored lepton asymmetry, the entropy production is so significant that the gravitinos do not account for the observed dark matter abundance.

L-balls gradually emit their charge even before the decay at $T_\mathrm{D}$ and a part of the lepton asymmetry emitted before the electroweak phase transition can be partially converted into a negative baryon asymmetry.
To avoid the overproduction of negative baryon asymmetry, we evaluate the abundance of the baryon asymmetry from the L-balls and set the limit on the parameters.
The emission rate of lepton number from an L-ball is determined by the evaporation of L-balls and the diffusion of the emitted charge.
Here, we assume 
\begin{align}
    T_R 
    \gtrsim
    T_* 
    &\sim
    M_s^{2/3} M_F^{1/3} Q^{-1/12}
    \sim
    M_s^{2/3} m_{3/2}^{1/3}
\nonumber\\
    &\sim
    3.68 \times 10^{2} ~\mathrm{GeV} \,
    \left( \frac{M_s}{ 10^4~\mathrm{GeV}} \right)^{2/3}
    \left( \frac{m_{3/2}}{0.5~\mathrm{GeV}} \right)^{1/3},
    \label{eq: TR lower bound for delayed}
\end{align}
where $T_*$ is the temperature when the diffusion is no longer the limiting factor of the evaporation rate, 
and $M_s$ is the sparticle mass.
Then, the total lepton number evaporated from a single L-ball $\Delta Q$ is given by~\cite{Kasuya:2014ofa}
\begin{align}
    \Delta Q
    \simeq
    2^{5/6} 9 \pi A^{2/3} \zeta^{-1/3} \sqrt{ \frac{10}{\tilde{g}_*} } 
    \frac{M_\mathrm{Pl}}{M_s^{2/3} M_F^{1/3}} Q^{1/12},
    \label{eq: dQ formula for delayed}
\end{align}
where $A \simeq 4$ is a coefficient of the diffusion constant and $\tilde{g}_*$ is a typical value of $g_*$ during the evaporation.
Since the lepton number evaporated from an L-ball before the electroweak phase transition $\Delta Q_\mathrm{EW}$ is the same order as $\Delta Q$, we assume $\Delta Q_\mathrm{EW} = \Delta Q$ in the following.
Since the evaporation is most efficient around the transition temperature $T_*$,
we use $\tilde{g}_* \sim 100$ as a typical value of $g_*$ at $T \sim T_*$ in the following.
Using $\Delta Q$, we can evaluate the baryon asymmetry originated from L-balls as
\begin{align}
    \eta_{B,Q}
    &=
    -\frac{8}{23}\frac{\Delta Q_\mathrm{EW}}{Q} \eta_L
\nonumber\\
    &=
    -\frac{3^{5/2} \pi^2 N_\ell^{1/2} \zeta^{1/6} A^{2/3}}{46 \beta^{37/24}}
    \sqrt{ \frac{10}{\tilde{g}_*} }
    \left( \frac{90}{\pi^2 g_*(T_\mathrm{D})} \right)^{1/4}
    \frac{M_\mathrm{Pl}^{3/2} m_{3/2}^{31/6}}{ M_s^{2/3} M_F^{6}}
\nonumber\\
    & \simeq 
    -6.83 \times 10^{-12} \,
    \left( \frac{N_\ell}{3} \right)^{1/2}
    \left( \frac{A}{4} \right)^{2/3}
    \left( \frac{\zeta}{2.5} \right)^{1/6}
    \left( \frac{\tilde{g}_*}{100} \right)^{-1/2}
    \left( \frac{g_*}{10.75} \right)^{-1/4}
\nonumber\\
    & \qquad \qquad \qquad \times 
    \left( \frac{M_s}{ 10^{4}~\mathrm{GeV}} \right)^{-2/3}
    \left( \frac{M_F}{5 \times 10^6~\mathrm{GeV}} \right)^{-6}
    \left( \frac{\beta}{6 \times 10^{-4}} \right)^{-37/24}
    \left( \frac{m_{3/2}}{0.5~\mathrm{GeV}} \right)^{31/6}
    .
\end{align}
In order not to spoil the success of other baryogenesis scenarios that explain the observed baryon asymmetry $\eta_{B,\mathrm{obs}} \sim 10^{-10}$, we require $|\eta_{B,Q}| \lesssim \eta_{B,\mathrm{obs}}$.

The generated lepton asymmetry and cosmological constraints are shown in Fig.~\ref{fig: lepton asymmetry and constraints}.
There, we use $M_s = 10^4$~GeV, $T_R = 10^5$~GeV, and $\varphi_\mathrm{osc} = 10^4 \varphi_\mathrm{eq}$.
The favored lepton asymmetry is shown by the red region, while the green and cyan regions are constrained by the requirement of $T_\mathrm{D} < 1$~MeV and $|\eta_{B,Q}| < 10^{-10}$.
In the blue region, the L-balls do not dominate the universe at the decay and our estimation of the lepton asymmetry is invalid.
As we can see, there is a parameter region including the benchmark point shown by the orange star where the favored lepton asymmetry is realized without violating the cosmological constraints.

Before closing this section, we briefly discuss the model parameters in the gauge-mediated SUSY breaking scenario.
In the minimal gauge-mediation model (e.g. see, \cite{Hamaguchi:2014sea}), $M_F$ and $m_{3/2}$ are given by
\begin{align}
    M_F  & = \frac{g^{1/2}}{4\pi}\sqrt{k\,F} ,\\
    m_{3/2} & = \frac{F}{\sqrt{3}M_\text{Pl}},
\end{align}
where $F$ is the vacuum expectation value of the SUSY breaking $F$-term, $g$ is the gauge coupling constant, and $k\,(\lesssim 1)$ is the coupling between the SUSY breaking field and the messenger field.
The Higgs mass measurement at LHC leads to a constraint on $F$ as $\sqrt{kF} \gtrsim 5\times 10^5$~GeV~\cite{Hamaguchi:2014sea}.
Thus, we obtain constraints on $M_F$ and $m_{3/2}$ as
\begin{align}
    M_F & \gtrsim 4\times 10^4 \sqrt{g} ~\text{GeV}, \\
    m_{3/2} & \gtrsim 6 \times 10^{-8} k^{-1}~\text{GeV}.
\end{align}
On the other hand, the gravitino mass cannot be larger than $\mathcal{O}(1)$~GeV to solve the flavor-changing neutral current problem. 
Our benchmark point in Fig.~\ref{fig: lepton asymmetry and constraints} satisfies these requirements.

\section{Summary and Discussion}
\label{sec: summary and discussion}

In this paper, motivated by the new determination of the primordial $^4$He abundance favoring a positive lepton asymmetry $\xi_e \sim 0.05$,
we have considered the formation, decay, and evaporation of L-balls and discussed the generated lepton asymmetry and cosmological constraints.
As a result, we have found that such a large lepton asymmetry can be realized using L-balls that decay after the electroweak phase transition and before BBN.
This contrasts with the fact that the maximum lepton asymmetry generated from right-handed neutrino decay is insufficient to realize $\xi_e \sim 0.05$.
With our choice of the model parameters, the L-balls dominate the universe at the decay and then the constraint from the gravitino problem is significantly relaxed.
Moreover, the evaporation of lepton asymmetry from the L-balls at high temperature can be sufficiently small, and thus, the negative baryon asymmetry converted from the evaporated lepton asymmetry does not require a fine-tuned baryogenesis.
The generated lepton asymmetry and relevant cosmological constraints are summarized in Fig.~\ref{fig: lepton asymmetry and constraints}.

While the constraints on $\xi_e$ from the observations other than BBN are consistent with both of $\xi_e = 0$ and $\xi_e = 0.05$~\cite{Popa:2008tb,Caramete:2013bua,Oldengott:2017tzj,Nunes:2017xon}, the new $^4$He observation favors $\xi_e = 0.05^{+0.03}_{-0.02}$, which is in moderate tension with no lepton asymmetry above the $2 \sigma$ level.
With an increase of the number of observed extremely metal-poor galaxies, the uncertainty of the primordial $^4$He abundance is expected to be reduced and so is that of $\xi_e$.
If nonzero $\xi_e$ is confirmed with a greater significance level in the future, our L-ball scenario becomes more attracting. 

The present model produces a large lepton asymmetry and a tiny opposite baryon asymmetry.
Thus, the observed positive baryon asymmetry should be provided by another mechanism.
Since there exist several flat directions with baryon charges in MSSM, it is possible for another flat direction to generate the baryon asymmetry. 

\begin{acknowledgments}
We would like to thank Shintaro Eijima, Masami Ouchi, and Akinori Matsumoto for useful discussions and comments.
We also thank the EMPRESS 3D members for their encouragement.
This work was supported by JSPS KAKENHI Grant Nos. 20H05851(M.K.), 21K03567(M.K.), JP20J20248 (K.M.) and World Premier International Research Center Initiative (WPI Initiative), MEXT, Japan (M.K., K.M.).
K.M. was supported by the Program of Excellence in Photon Science.
\end{acknowledgments}

\small
\bibliographystyle{JHEP}
\bibliography{Ref}

\providecommand{\href}[2]{#2}\begingroup\raggedright\begin{thebibliography}{10}

\bibitem{Matsumoto:2022tlr}
A.~Matsumoto et~al., \emph{{EMPRESS. VIII. A New Determination of Primordial He
  Abundance with Extremely Metal-Poor Galaxies: A Suggestion of the Lepton
  Asymmetry and Implications for the Hubble Tension}},
  \href{https://arxiv.org/abs/2203.09617}{{\ttfamily 2203.09617}}.

\bibitem{Hsyu:2020uqb}
T.~Hsyu, R.~J. Cooke, J.~X. Prochaska and M.~Bolte, \emph{{The PHLEK Survey: A
  New Determination of the Primordial Helium Abundance}},
  \href{https://doi.org/10.3847/1538-4357/ab91af}{\emph{Astrophys. J.}
  {\bfseries 896} (2020) 77},
  [\href{https://arxiv.org/abs/2005.12290}{{\ttfamily 2005.12290}}].

\bibitem{Aver:2015iza}
E.~Aver, K.~A. Olive and E.~D. Skillman, \emph{{The effects of He I
  \ensuremath{\lambda}10830 on helium abundance determinations}},
  \href{https://doi.org/10.1088/1475-7516/2015/07/011}{\emph{JCAP} {\bfseries
  07} (2015) 011}, [\href{https://arxiv.org/abs/1503.08146}{{\ttfamily
  1503.08146}}].

\bibitem{Izotov:2014fga}
Y.~I. Izotov, T.~X. Thuan and N.~G. Guseva, \emph{{A new determination of the
  primordial He abundance using the He i $\lambda$10830 \r{A} emission line:
  cosmological implications}},
  \href{https://doi.org/10.1093/mnras/stu1771}{\emph{Mon. Not. Roy. Astron.
  Soc.} {\bfseries 445} (2014) 778--793},
  [\href{https://arxiv.org/abs/1408.6953}{{\ttfamily 1408.6953}}].

\bibitem{Cooke:2017cwo}
R.~J. Cooke, M.~Pettini and C.~C. Steidel, \emph{{One Percent Determination of
  the Primordial Deuterium Abundance}},
  \href{https://doi.org/10.3847/1538-4357/aaab53}{\emph{Astrophys. J.}
  {\bfseries 855} (2018) 102},
  [\href{https://arxiv.org/abs/1710.11129}{{\ttfamily 1710.11129}}].

\bibitem{Planck:2018vyg}
{\scshape Planck} collaboration, N.~Aghanim et~al., \emph{{Planck 2018 results.
  VI. Cosmological parameters}},
  \href{https://doi.org/10.1051/0004-6361/201833910}{\emph{Astron. Astrophys.}
  {\bfseries 641} (2020) A6},
  [\href{https://arxiv.org/abs/1807.06209}{{\ttfamily 1807.06209}}].

\bibitem{Seto:2021tad}
O.~Seto and Y.~Toda, \emph{{Hubble tension in lepton asymmetric cosmology with
  an extra radiation}},
  \href{https://doi.org/10.1103/PhysRevD.104.063019}{\emph{Phys. Rev. D}
  {\bfseries 104} (2021) 063019},
  [\href{https://arxiv.org/abs/2104.04381}{{\ttfamily 2104.04381}}].

\bibitem{Fukugita:1986hr}
M.~Fukugita and T.~Yanagida, \emph{{Baryogenesis Without Grand Unification}},
  \href{https://doi.org/10.1016/0370-2693(86)91126-3}{\emph{Phys. Lett. B}
  {\bfseries 174} (1986) 45--47}.

\bibitem{Affleck:1984fy}
I.~Affleck and M.~Dine, \emph{{A New Mechanism for Baryogenesis}},
  \href{https://doi.org/10.1016/0550-3213(85)90021-5}{\emph{Nucl. Phys. B}
  {\bfseries 249} (1985) 361--380}.

\bibitem{Dine:1995kz}
M.~Dine, L.~Randall and S.~D. Thomas, \emph{{Baryogenesis from flat directions
  of the supersymmetric standard model}},
  \href{https://doi.org/10.1016/0550-3213(95)00538-2}{\emph{Nucl. Phys. B}
  {\bfseries 458} (1996) 291--326},
  [\href{https://arxiv.org/abs/hep-ph/9507453}{{\ttfamily hep-ph/9507453}}].

\bibitem{Boyarsky:2009ix}
A.~Boyarsky, O.~Ruchayskiy and M.~Shaposhnikov, \emph{{The Role of sterile
  neutrinos in cosmology and astrophysics}},
  \href{https://doi.org/10.1146/annurev.nucl.010909.083654}{\emph{Ann. Rev.
  Nucl. Part. Sci.} {\bfseries 59} (2009) 191--214},
  [\href{https://arxiv.org/abs/0901.0011}{{\ttfamily 0901.0011}}].

\bibitem{Coleman:1985ki}
S.~R. Coleman, \emph{{Q Balls}},
  \href{https://doi.org/10.1016/0550-3213(86)90520-1}{\emph{Nucl. Phys. B}
  {\bfseries 262} (1985) 263}.

\bibitem{Kusenko:1997si}
A.~Kusenko and M.~E. Shaposhnikov, \emph{{Supersymmetric Q balls as dark
  matter}}, \href{https://doi.org/10.1016/S0370-2693(97)01375-0}{\emph{Phys.
  Lett. B} {\bfseries 418} (1998) 46--54},
  [\href{https://arxiv.org/abs/hep-ph/9709492}{{\ttfamily hep-ph/9709492}}].

\bibitem{Enqvist:1997si}
K.~Enqvist and J.~McDonald, \emph{{Q balls and baryogenesis in the MSSM}},
  \href{https://doi.org/10.1016/S0370-2693(98)00271-8}{\emph{Phys. Lett. B}
  {\bfseries 425} (1998) 309--321},
  [\href{https://arxiv.org/abs/hep-ph/9711514}{{\ttfamily hep-ph/9711514}}].

\bibitem{Kasuya:1999wu}
S.~Kasuya and M.~Kawasaki, \emph{{Q ball formation through Affleck-Dine
  mechanism}}, \href{https://doi.org/10.1103/PhysRevD.61.041301}{\emph{Phys.
  Rev. D} {\bfseries 61} (2000) 041301},
  [\href{https://arxiv.org/abs/hep-ph/9909509}{{\ttfamily hep-ph/9909509}}].

\bibitem{Kawasaki:2002hq}
M.~Kawasaki, F.~Takahashi and M.~Yamaguchi, \emph{{Large lepton asymmetry from
  Q balls}}, \href{https://doi.org/10.1103/PhysRevD.66.043516}{\emph{Phys. Rev.
  D} {\bfseries 66} (2002) 043516},
  [\href{https://arxiv.org/abs/hep-ph/0205101}{{\ttfamily hep-ph/0205101}}].

\bibitem{Gelmini:2020ekg}
G.~B. Gelmini, M.~Kawasaki, A.~Kusenko, K.~Murai and V.~Takhistov, \emph{{Big
  Bang Nucleosynthesis constraints on sterile neutrino and lepton asymmetry of
  the Universe}},
  \href{https://doi.org/10.1088/1475-7516/2020/09/051}{\emph{JCAP} {\bfseries
  09} (2020) 051}, [\href{https://arxiv.org/abs/2005.06721}{{\ttfamily
  2005.06721}}].

\bibitem{Hisano:2001dr}
J.~Hisano, M.~M. Nojiri and N.~Okada, \emph{{The Fate of the B ball}},
  \href{https://doi.org/10.1103/PhysRevD.64.023511}{\emph{Phys. Rev. D}
  {\bfseries 64} (2001) 023511},
  [\href{https://arxiv.org/abs/hep-ph/0102045}{{\ttfamily hep-ph/0102045}}].

\bibitem{Kawasaki:2012gk}
M.~Kawasaki and M.~Yamada, \emph{{$Q$ ball Decay Rates into Gravitinos and
  Quarks}}, \href{https://doi.org/10.1103/PhysRevD.87.023517}{\emph{Phys. Rev.
  D} {\bfseries 87} (2013) 023517},
  [\href{https://arxiv.org/abs/1209.5781}{{\ttfamily 1209.5781}}].

\bibitem{Kasuya:2014ofa}
S.~Kasuya and M.~Kawasaki, \emph{{Baryogenesis from the gauge-mediation type
  Q-ball and the new type of Q-ball as the dark matter}},
  \href{https://doi.org/10.1103/PhysRevD.89.103534}{\emph{Phys. Rev. D}
  {\bfseries 89} (2014) 103534},
  [\href{https://arxiv.org/abs/1402.4546}{{\ttfamily 1402.4546}}].

\bibitem{Gherghetta:1995dv}
T.~Gherghetta, C.~F. Kolda and S.~P. Martin, \emph{{Flat directions in the
  scalar potential of the supersymmetric standard model}},
  \href{https://doi.org/10.1016/0550-3213(96)00095-8}{\emph{Nucl. Phys. B}
  {\bfseries 468} (1996) 37--58},
  [\href{https://arxiv.org/abs/hep-ph/9510370}{{\ttfamily hep-ph/9510370}}].

\bibitem{Dine:2003ax}
M.~Dine and A.~Kusenko, \emph{{The Origin of the matter - antimatter
  asymmetry}}, \href{https://doi.org/10.1103/RevModPhys.76.1}{\emph{Rev. Mod.
  Phys.} {\bfseries 76} (2003) 1},
  [\href{https://arxiv.org/abs/hep-ph/0303065}{{\ttfamily hep-ph/0303065}}].

\bibitem{deGouvea:1997afu}
A.~de~Gouvea, T.~Moroi and H.~Murayama, \emph{{Cosmology of supersymmetric
  models with low-energy gauge mediation}},
  \href{https://doi.org/10.1103/PhysRevD.56.1281}{\emph{Phys. Rev. D}
  {\bfseries 56} (1997) 1281--1299},
  [\href{https://arxiv.org/abs/hep-ph/9701244}{{\ttfamily hep-ph/9701244}}].

\bibitem{Kusenko:1997zq}
A.~Kusenko, \emph{{Solitons in the supersymmetric extensions of the standard
  model}}, \href{https://doi.org/10.1016/S0370-2693(97)00584-4}{\emph{Phys.
  Lett. B} {\bfseries 405} (1997) 108},
  [\href{https://arxiv.org/abs/hep-ph/9704273}{{\ttfamily hep-ph/9704273}}].

\bibitem{Dvali:1997qv}
G.~Dvali, A.~Kusenko and M.~E. Shaposhnikov, \emph{{New physics in a nutshell,
  or Q ball as a power plant}},
  \href{https://doi.org/10.1016/S0370-2693(97)01378-6}{\emph{Phys. Lett. B}
  {\bfseries 417} (1998) 99--106},
  [\href{https://arxiv.org/abs/hep-ph/9707423}{{\ttfamily hep-ph/9707423}}].

\bibitem{Kasuya:2001hg}
S.~Kasuya and M.~Kawasaki, \emph{{Q ball formation: Obstacle to Affleck-Dine
  baryogenesis in the gauge mediated SUSY breaking?}},
  \href{https://doi.org/10.1103/PhysRevD.64.123515}{\emph{Phys. Rev. D}
  {\bfseries 64} (2001) 123515},
  [\href{https://arxiv.org/abs/hep-ph/0106119}{{\ttfamily hep-ph/0106119}}].

\bibitem{Cohen:1986ct}
A.~G. Cohen, S.~R. Coleman, H.~Georgi and A.~Manohar, \emph{{The Evaporation of
  $Q$ Balls}}, \href{https://doi.org/10.1016/0550-3213(86)90004-0}{\emph{Nucl.
  Phys. B} {\bfseries 272} (1986) 301--321}.

\bibitem{Kawasaki:2017bqm}
M.~Kawasaki, K.~Kohri, T.~Moroi and Y.~Takaesu, \emph{{Revisiting Big-Bang
  Nucleosynthesis Constraints on Long-Lived Decaying Particles}},
  \href{https://doi.org/10.1103/PhysRevD.97.023502}{\emph{Phys. Rev. D}
  {\bfseries 97} (2018) 023502},
  [\href{https://arxiv.org/abs/1709.01211}{{\ttfamily 1709.01211}}].

\bibitem{Hamaguchi:2014sea}
K.~Hamaguchi, M.~Ibe, T.~T. Yanagida and N.~Yokozaki, \emph{{Testing the
  Minimal Direct Gauge Mediation at the LHC}},
  \href{https://doi.org/10.1103/PhysRevD.90.015027}{\emph{Phys. Rev. D}
  {\bfseries 90} (2014) 015027},
  [\href{https://arxiv.org/abs/1403.1398}{{\ttfamily 1403.1398}}].

\bibitem{Popa:2008tb}
L.~A. Popa and A.~Vasile, \emph{{WMAP 5-year constraints on lepton asymmetry
  and radiation energy density: Implications for Planck}},
  \href{https://doi.org/10.1088/1475-7516/2008/06/028}{\emph{JCAP} {\bfseries
  06} (2008) 028}, [\href{https://arxiv.org/abs/0804.2971}{{\ttfamily
  0804.2971}}].

\bibitem{Caramete:2013bua}
A.~Caramete and L.~A. Popa, \emph{{Cosmological evidence for leptonic asymmetry
  after Planck}},
  \href{https://doi.org/10.1088/1475-7516/2014/02/012}{\emph{JCAP} {\bfseries
  02} (2014) 012}, [\href{https://arxiv.org/abs/1311.3856}{{\ttfamily
  1311.3856}}].

\bibitem{Oldengott:2017tzj}
I.~M. Oldengott and D.~J. Schwarz, \emph{{Improved constraints on lepton
  asymmetry from the cosmic microwave background}},
  \href{https://doi.org/10.1209/0295-5075/119/29001}{\emph{EPL} {\bfseries 119}
  (2017) 29001}, [\href{https://arxiv.org/abs/1706.01705}{{\ttfamily
  1706.01705}}].

\bibitem{Nunes:2017xon}
R.~C. Nunes and A.~Bonilla, \emph{{Probing the properties of relic neutrinos
  using the cosmic microwave background, the Hubble Space Telescope and galaxy
  clusters}}, \href{https://doi.org/10.1093/mnras/stx2661}{\emph{Mon. Not. Roy.
  Astron. Soc.} {\bfseries 473} (2018) 4404--4409},
  [\href{https://arxiv.org/abs/1710.10264}{{\ttfamily 1710.10264}}].

\end{thebibliography}\endgroup

\end{document}